\documentclass{amsart}
\usepackage{amssymb}
\usepackage{amsmath}
\usepackage{amsfonts}
\usepackage{graphicx}
\usepackage{epstopdf}

\usepackage{young}
\usepackage[vcentermath]{youngtab}
\usepackage{ytableau}

\usepackage{tikz}
\usepackage{float}
\usepackage{placeins}

\newcommand{\beq}{\begin{equation}}
\newcommand{\eeq}{\end{equation}}

\begin{document}

\begin{center}
{\Large\bf $N \leftrightarrow -N$ duality of SU(N) for stable sequences of representations
}\\
\vspace*{1 cm}
{\large  }
\vspace*{0.5 cm}

{\large  R.L.Mkrtchyan 
}

\vspace*{0.5 cm}

{\small\it Yerevan Physics Institute, 2 Alikhanian Br. Str., 0036 Yerevan, Armenia}

{\small\it E-mail: mrl55@list.ru}

\vspace*{0.5 cm}

\end{center}

 {\bf Abstract.} We generalize $N \leftrightarrow -N$ duality of dimension formulae of $SU(N)$ representations on a (class of) representations with $N$-dependent Young diagrams (which include adjoint representation), and on eigenvalues of Casimir operator for that representations. We discuss the consequences for the hypothesis of universal decomposition of powers of adjoint representation into Casimir subspaces.

\section{Introduction}

There exists a self-duality of $SU(N)$ theory, and similar duality  $SO(N) \leftrightarrow Sp(N)$, under appropriately defined $N \leftrightarrow -N$ transformation, established, in various contexts, in \cite{RP},  \cite{King}, \cite{PS}, \cite{Cvit,Cvitbook}, \cite{Mkr}, and other papers. 

For example,   \cite{King}, dimensions of irreducible representations of $SU(N)$ with Young diagram $Y$ are covariant under $N \leftrightarrow -N$    

\begin{eqnarray} \label{sun}
	dim_{su}(Y,N)=(-1)^{Area(Y)} dim_{su}(Y^T,-N)
\end{eqnarray}
where $Y^T$ denotes the transposed $Y$ diagram, $Area(Y)$ is the number of boxes in $Y$. Similarly,  dimensions of tensor representations of $so(N)$ and $sp(N)$ ($N$ even) are connected under $N\leftrightarrow -N$ 

\begin{eqnarray}\label{sosp}
	dim_{so}(Y,N)=(-1)^{Area(Y)} dim_{sp}(Y^T,-N)
\end{eqnarray}

One general explanation of these dualities is the interpretation \cite{PS} of the space with negative dimensionality  as the space with Grassmann coordinates. This leads to the switching of symmetrization and antisymmetrization in comparison with formulae in space with commuting coordinates ("negative dimensions theorems" \cite{Cvitbook}). 

Another general approach to this phenomenon is Vogel's universality \cite{V0}. Vogel's approach, originated from the study of knot theory's weight systems of Jacoby diagrams, leads to the parametrization of all simple Lie algebras, and some their generalizations, in terms of homogeneous coordinates $\alpha, \beta, \gamma$ of projective plane. They are relevant up to rescaling and permutations, and parameterize algebras by Vogel's table. Particularly, for $su(N), so(N), sp(N)$ one can take  $(\alpha, \beta, \gamma)=(-2,2,N), (-2,4,N-4), (-2,1,N/2+2)$, repectively. It is easy to see that under  $N \leftrightarrow -N$ Vogel's parameters of $su$ can be transformed back to themselves with appropriate permutation and rescaling, and parameters of $so$ and $sp$ transformes one into another. 

However, these general understandings do not lead immediately, or even in any regular way, to formulae like (\ref{sun}), (\ref{sosp}), with all their details. 

The aim of the present work is to extend $N\leftrightarrow -N$ self-duality of $su(N)$ algebra for some other class of representations, which are not covered by (\ref{sun}), and seems not to be presented in literature, namely those with specific $N$-dependence of Young diagrams. 

To illustrate the problem let's first consider the proof of duality \ref{sun}.  It is simple, due to hook formula for dimensions of $su(N)$.  Consider, e.g. the dimension of irrep with Young diagram 

\begin{eqnarray}
	\ydiagram{3,3,1}
\end{eqnarray}

According to hook formula the dimension of corresponding irrep of $su(N)$ is given by

\begin{eqnarray}
dim=\frac{Num}{Den}
\end{eqnarray}
where both $Num$ and $Den$ are the products of numbers in the boxes of given Young diagram. For numerator that is 

\begin{eqnarray}
	\ytableausetup{mathmode, boxsize=2em}
	\begin{ytableau}
		\scriptstyle N& 	\scriptstyle N+1 &	\scriptstyle N+2 \\
			\scriptstyle N-1 &	\scriptstyle N   & 	\scriptstyle N+1 \\
			\scriptstyle N-2 
	\end{ytableau}
\end{eqnarray}
i.e. one put $N$ in left upper box, and in any other box, located $i$ steps to the right and $j$ steps lower, put $N+i-j$.

The denominator is the product of hooks lengths, i.e. in each box one put the length of its hook, i.e. the number of boxes, including itself, in the row to the right of it, and in column below it. For our example that is:

\begin{eqnarray}
 \ytableaushort{
	532,
	421,
	1
}
\end{eqnarray}

Denominator doesn't change  under $N \leftrightarrow -N$, and is the same for transposed diagram. It is easy to understand, that numerator under  $N \leftrightarrow -N$ becomes the numerator of the transposed diagram, up the minus sign for each box, so totally one obtain minus in the power of area of diagram. 

This example make it clear the theorem (\ref{sun}) for an arbitrary Young diagram.

However, consider the Young diagram of an adjoint representation of $su(N)$: 

\begin{eqnarray}
 \begin{ytableau}
	~ & \\
	~  \\
	\none[\vdots] \\
	~\\ 
\end{ytableau}
\end{eqnarray}

where the number of squares in the first column is $N-1$. Dimension can be easily calculated by hook formula: 

\begin{eqnarray}
	Num =  \begin{ytableau}
		\scriptstyle N & 	\scriptstyle N+1 \\
			\scriptstyle N -1 \\
	\none[\vdots] \\
			\scriptstyle 2
	\end{ytableau}, \,\,
	Den =  \begin{ytableau}
		\scriptstyle N & 	\scriptstyle 1 \\
		\scriptstyle N -2 \\
		\none[\vdots] \\
		\scriptstyle 1\\ 
	\end{ytableau} \\
dim=	\frac{Num}{Den}= N^2-1
\end{eqnarray}

If we try to literally  apply (\ref{sun}) to this diagram, we will obtain transposed diagram 

\begin{eqnarray}
	\begin{ytableau}
		~ & ~ & \none[\cdots] &\\
		~\\ 
	\end{ytableau}
\end{eqnarray}

with dimension of corresponding representation 

\begin{eqnarray}
	\frac{(2N-2)!(N-1)}{(N-2)!N!}
\end{eqnarray}
which is not $N^2-1$ of adjoint. 

The problem of course is in that (\ref{sun}) implies that parameters of diagram $Y$ are $N$-independent, while in this example the height of the first column (in the initial diagram) or in the first row (in the transposed diagram) are $N$-dependent and introduce additional $N$-dependence in the dimension of corresponding representation. 

We shall  extend below the scope of $N\leftrightarrow -N$ theorem for $su(N)$ in such a way that it will cover the case of adjoint representation, and similar representations, denoted $D(\lambda,\tau)$  and which is a particular case of  {\it stable sequences} of representations, see below. One should first describe the notion of $N$-dependence.

\section{Stable sequences}

When we speak about the dependence of $N$ of dimension, or other quantity, one should define the sequence of representations  for different $N$.  Let we characterize representation of $su(M)$, at some $M$, by Dynkin labels $(\lambda_1, \lambda_2, ..., \lambda_{M-1})$, then one natural way to extend this representation for higher ranks is to add zeros from the right, i.e. for an arbitrary $N>M$ consider representation with labels $(\lambda_1, \lambda_2, ..., \lambda_{M-1},0,...,0)$. Then representations will "maintain their names": fundamental vector representation, symmetric tensor, etc. will remain so. Another characterization of this sequence is that we keep their Young diagram. However, representation $(1,0,...,0,1)$ has a special name - adjoint representation, and to maintain it one should add zeros between $1$-s, keeping them on the first and last positions. In this case Young diagram is changing, first column is rising according to $N$.

Evidently, one can extend a given set of Dynkin labels in various ways while preserving the non-zero entries unchanged, but adding zeros with some rules in between of non-zero labels, i.e. lengthening some columns as $N$ is rising.  Of course, the choice depends from the specific problem under study. In our case one can suggest the following principle: dimensions of representations should remain polynomial. Then the natural rules seems to be just described ones, only (one can add also the third rule - adding zeros at the left end of Dynkin labels, which is essentially the same as doing so on the right end, due to automorphism of Dynkin diagram). We have no much to say further on this issue, and concentrate on the rule of adding zeros in the middle of row of Dynkin labels, otherwise leaving them the same. In \cite{MRL25} such sequences were called the stable sequences of representations. 

So, we consider sequences of representations with Dynkin labels of the type 

\begin{eqnarray} \label{stabsec}
	(\lambda_,...,\lambda_k,0,...,0,\tau_k,...,\tau_1)
\end{eqnarray}
The possibly non-zero labels at the begiining and the end of the row of Dynkin labels remain the same as $N$ changes, only the number of zeros in between changes. 

Such representations are denoted  $D(\lambda,\tau)$.  

In the next sections we present the generalization  of $N \leftrightarrow -N$ theorem (\ref{sun}) for these representations.

\section{Example} 

Consider stable sequence with only non-zero labels $\lambda_1=2, \lambda_2=1, \lambda_3=1$, and $\tau_1=1, \tau_3=1$. 

Corresponding Young diagram is shown on the following picture: 

\begin{eqnarray} \label{Example}
 \begin{ytableau}
	a& 	a &\lambda&\lambda&\lambda&\lambda \\
   a&a&\lambda&\lambda\\
    a&a&\lambda\\
    b&b \\
	\none[\vdots] &\none[\vdots]\\
	b&b\\
b&\none[\bar{\tau}] \\
b &\none[\bar{\tau}]\\
\none[\bar{\tau}]&\none[\bar{\tau}]
\end{ytableau}
\end{eqnarray}

The Young diagram itself consists from boxes labeled $a,b$ and $\lambda$. We also show the complement of columns, corresponding to non-zero $\tau$-s, labeling corresponding invisible boxes by $\bar{\tau}$. We draw that complement ("inverted Young diagram") below, denoting  it (and other inverted diagrams) by letter with bar, in this case $\bar{\tau}$:

\begin{eqnarray}
	\ytableausetup{smalltableaux}
	\bar{\tau}&=&\ydiagram{1+1,1+1,2}
	\ytableausetup{nosmalltableaux}
\end{eqnarray}

For this inverted Young diagram we put in correspondence the usual Young diagram $\tau$, which is a reflection of inverted one w.r.t. the line with $45^\circ$ slope (line y=x in the usual parameterization of plane):

\begin{eqnarray}
	\ytableausetup{smalltableaux}
 \tau=\ydiagram{3,1} 
	\ytableausetup{nosmalltableaux}
\end{eqnarray}

We also introduce the Young diagram $\lambda$, which is the diagram, corresponding to the first three non-zero Dynkin labels, and which already is indicated on diagram (\ref{Example}) above as subdiagram of boxes with $\lambda$ inside:
\begin{eqnarray}
	\ytableausetup{smalltableaux}
	\lambda&=&\ydiagram{4,2,1}
\end{eqnarray}

and also introduce its inverted version

\begin{eqnarray}
	\ytableausetup{smalltableaux}
	\bar{\lambda}&=&\ydiagram{2+1,2+1,1+2,3}
		\ytableausetup{nosmalltableaux}
\end{eqnarray}

According to (\ref{Example}) our diagram consist from three parts, $a,b$, and $\lambda$, with boxes denoted by the corresponding letters. 

With this data, we construct the new diagram below. Dimension of corresponding irrep, as we shall show, is connected with dimension of initial diagram, by $N \leftrightarrow -N$ transformation, up to the sign. 

New diagram is:

\begin{eqnarray} \label{Example1}
	\begin{ytableau}
		a'& 	a' & a' &\tau&\tau&\tau \\
		a'&a'&a'&\tau\\
		b'&b'&b'\\
		\none[\vdots] &\none[\vdots]&\none[\vdots]\\
		b'&b'&b'\\
		b'&b'&\none[\bar{\lambda}] \\
		b'&b'&\none[\bar{\lambda}] \\
		b' &\none[\bar{\lambda}] &\none[\bar{\lambda}]  \\
	\none[\bar{\lambda}]&\none[\bar{\lambda}]&\none[\bar{\lambda}]
	\end{ytableau}
\end{eqnarray}

This new diagram also consists from three building bloks: $a', b'$ and $\tau$. $a'$ is transposition of $a$, $b'$ with $a'$ constitute complement of $\bar{\lambda}$ to columns of hight $N$. Let us calculate and compare the contributions into dimension of diagram of the pairs $a$ and  $a'$,  $b'$ and $\lambda$, $b$ and $\tau$. 

We should fill corresponding boxes with multipliers according to the hook formula. 
Consider pair  $b$ and $\tau$, we have for $\tau$:

\begin{eqnarray} \label{tau}
	\begin{ytableau}
	\scriptstyle	N+3&	\scriptstyle N+4& 	\scriptstyle N+5 \\
		\scriptstyle N+2
	\end{ytableau} 
	/
	\begin{ytableau}
		\scriptstyle	4&	\scriptstyle 2& 	\scriptstyle 1 \\
		\scriptstyle 1
	\end{ytableau}
\end{eqnarray}

For $b$ we note that in left upper box there is $N-3$, and dimension of this contribution can be understood to be that of the transposed $\tau$ diagram $\tau^T$ for algebra $su(N-3)$, due to $Z_2$ automorphism of $su$ algebras, i.e. 

\begin{eqnarray} 
	\begin{ytableau}
		\scriptstyle	N-3&\scriptstyle N-2 \\
		\scriptstyle N-4 \\
		\scriptstyle N-5 
	\end{ytableau} 
	/
	\begin{ytableau}
		\scriptstyle	4&	\scriptstyle 2& 	\scriptstyle 1 \\
		\scriptstyle 1
	\end{ytableau}
\end{eqnarray}

Immediate observation is that this two contributions are connected by $N\leftrightarrow -N$, up to sign $(-1)^{Area(\tau)}$. 

Evidently, due to the symmetry between $\lambda$ and $\tau$, similar relation is between contributions of $b'$ and $\lambda$, i.e. they are connected by the  $N\leftrightarrow -N$,  up to the sign $(-1)^{Area(\lambda)}$.

Finally, compare contributions of $a$ and $a'$: 

\begin{eqnarray} \label{Example-a}
a: \,\,\,	\begin{ytableau}
		\scriptstyle N  & 	\scriptstyle N +1 \\
			\scriptstyle N -1&	\scriptstyle N\\
			\scriptstyle N -2 &	\scriptstyle N -1
	\end{ytableau}
	/
		\begin{ytableau}
		\scriptstyle N+4 & 	\scriptstyle N +1 \\
		\scriptstyle N +1&	\scriptstyle N-2\\
		\scriptstyle N -1&	\scriptstyle N -4
	\end{ytableau}
\end{eqnarray}

\begin{eqnarray} 
	a': \,\,\,	\begin{ytableau}
		\scriptstyle N  & 	\scriptstyle N +1 &		\scriptstyle N +2 \\
			\scriptstyle N -1&\scriptstyle N  &	\scriptstyle N +1
	\end{ytableau}
	/
	\begin{ytableau}
		\scriptstyle N+4 & 	\scriptstyle N +2 &		\scriptstyle N -1\\
			\scriptstyle N+1 & 	\scriptstyle N -1&	\scriptstyle N -4
	\end{ytableau}
\end{eqnarray}
These expressions are going one into another under $N \leftrightarrow -N$ without additional sign. 

So, altogether we obtain that dimensions of two diagrams (\ref{Example}) and (\ref{Example1}) are going one into another under  $N \leftrightarrow -N$ up to the sign $(-1)^{Area(\lambda)+Area(\tau)}$, which for these diagrams is equal to $(-1)$.

\section{Duality of dimensions - general case }

Let us denote dimension of sequence of representations (\ref{stabsec}) $D(\lambda,\tau)$ of $su(N)$ by 

\begin{eqnarray}
	dim(D(\lambda,\tau),N)
\end{eqnarray}

The general case is naturally suggested by the previous example to be 

{\bf Proposition 1}
\begin{eqnarray} \label{Main}
	dim(D(\lambda,\tau),N)=(-1)^{Area(\lambda)+Area(\tau)} dim(D(\tau, \lambda),-N)
\end{eqnarray}

Note that in this case dimension is polynomial over $N$, so changing the sign is clear. Note also that according to the $Z_2$ automorphism of Dynkin diagram of $su(N)$ we also have the equation

\begin{eqnarray}
	dim(D(\lambda,\tau),N)=dim(D(\tau^T,\lambda^T),N)
\end{eqnarray}

The proof of the general case goes along the same lines as above. We divide the Young diagram of representation $D(\lambda,\tau)$ into three pieces: one is Young diagram  $\lambda_i$, second is rectangular $a$ at the left of $\lambda$, and remaining part below $a$, denote it $b$, which is rectangular minus $\bar{\tau}$.  Again, the diagram of representation $D(\tau, \lambda)$ also is divided into three pieces, one is diagram $\tau$, then rectangular diagram $a'$ which is transposed $a$, and remaining part below $a'$, which is rectangular minus $\bar{\lambda}$. The contribution of the $b$  part of the initial diagram is equal to the dimension of representation with Young diagram $\tau^T$ for $su(N-\Lambda)$ algebra, where $\Lambda$ is the number of rows  of diagram $\lambda$. According to the $N \leftrightarrow -N$ duality of $N$ independent diagrams, this is equal to the dimension of representation of $su(N-\Lambda)$ with diagram $\tau$, up to the sign $(-1)^{Area(\tau)}$. This is equal to the contribution of $\tau$ part of the transformed diagram,  up to the sign $(-1)^{Area(\tau)}$. Similarly one prove the equality of the contributions of $\lambda$ (of initial) and $b'$ (of transformed) parts, up to the sign $(-1)^{Area(\lambda)}$. Finally, $a$ and $a'$ parts are connected by  $N \leftrightarrow -N$ transformation without additional sign. Indeed, their contributions are equal (each of them), to the ratio of numerator and denominator. Numerators of $a$ and $a'$ evidently are connected by $N \leftrightarrow -N$ transformation, up to the sign $(-1)^{Area(a)}=(-1)^{Area(a')}$. Denominators are the product over boxes of hooks lengths $h(i,j)$, with $i=1,2,...T;j=1,2,...,\Lambda$, where $T$ is the number of rows of diagram $\tau$. I.e. $a$ is the rectangular with $\Lambda$ rows and $T$ columns. If we denote $l_i$ the length of $i$-th row of diagram $\lambda$, and $t_i$ the same for $\tau$, then denominators for $a$ and $a'$ are, respectively: 
 
 \begin{eqnarray}
a: \,\,\, \prod_{i=1}^{T} \prod_{j=1}^{\Lambda} (T+1-i+l_i+N-j-t_j) \\
a': \,\,\,  \prod_{i=1}^{T} \prod_{j=1}^{\Lambda} (\Lambda+1-j+t_j+N-i-l_i) 
 \end{eqnarray}

In the second expression (for $a'$) we change $N \leftrightarrow -N$ and also make a change of variables $j \rightarrow \Lambda -j+1, i \rightarrow T -i+1$, and obtain

\begin{eqnarray}
a': \,\,\,   (-1)^{Area(a)}	\prod_{i=1}^{T} \prod_{j=1}^{\Lambda} (T+1-i+l_i+N-j-t_j) 
\end{eqnarray}
i.e. we obtain answer for $a$, up to the sign.  The signs for numerator and denominator cancel each other, and this finishs the proof of (\ref{Main}).

\section{Duality of Casimir's eigenvalues}

Besides dimensions, there are other interesting and important characteristics of irreducible representations. Among these are eigenvalues of Casimir operators. In \cite{MV11} the duality of generating functions of spectra of Casimirs \cite{pp} of all orders, on all representations, was established for usual sequences of representation, in \cite{MSV}  the universality of eigenvalues was established for all orders Casimirs on adjoint representation, similar universality is established for second order Casimir's eigenvalues on some series of universal representations in \cite{AM19}.  These features actually depend on definition of Casimir operators, as discussed in \cite{MSV}, since higher order Casimirs are defined up to the combination of lower order Casimirs. It seems that choice of \cite{MSV} is relevant to our considerations, also, however, for that choice spectra of Casimirs is not calculated (for an arbitrary representation, as we need). So we restrict considerations by the original Casimir (or in other wording Casimir of second order), where ambiguity is absent.  

If $X^a$ are generators of simple Lie algebra in some representation, and $g_{ab}$ an invariant metric, then  the  Casimir operator can be defined as

\begin{eqnarray}
	C=g_{ab}X^aX^b
\end{eqnarray}

Invariant metrics $g_{ab}$ in simple Lie algebras are unique up to the normalization, which we choose to be the minimal one, where the squares of the long roots of the algebra are equal to 2. 

If $\lambda$ is the highest weight vector, then Casimir's eigenvalue on the corresponding representation is (\cite{BR,Difr,GG}):

\begin{eqnarray}\label{Ceig}
	C=(\lambda,\lambda)+2(\lambda,\rho)
\end{eqnarray}
where $\rho$ is the Weyl vector in the root space, the half-sum of all positive roots of a given algebra, and the scalar product is with (the inverse of) the abovementioned metric.

We state that eigenvalues of $C$ on representations $D(\lambda,\tau)$ satisfy the duality relation under $N \leftrightarrow -N$:

{\bf Proposition 2}
\begin{eqnarray} \label{dY}
	C(D(\lambda, \tau), N)=-C(D(\lambda^T, \tau^T), -N)
\end{eqnarray}

Proof is as follows. The eigenvalue of this Casimir on $D(\lambda,\tau)$ is given in \cite{MRL25}:

\begin{eqnarray}
	C(D(\lambda, \tau), N)=N \sum_{i=1}^{k}  \left(  i\lambda_i+  i\tau_i   \right) +    \\
	\sum_{i,j=1}^{k} min(i,j) \lambda_i \lambda_j+\sum_{i,j=1}^{k} min(i,j) \tau_i \tau_j  - \sum_{i=1}^{k} \left(  i^2\lambda_i+i^2\tau_i   \right) - \\
	\frac{1}{N} \left( \sum_{i=1}^{k} ( i\lambda_i -i\tau_i ) \right)^2
\end{eqnarray}

The eigenvalues of Casimir in the case when all $\tau_i$ are zero, are 

\begin{eqnarray} \label{Y}
	C(D(\lambda, 0), N)=	\\
	N \sum_{i=1}^{k}   i\lambda_i +    
	\sum_{i,j=1}^{k} min(i,j) \lambda_i \lambda_j  - \sum_{i=1}^{k}   i^2\lambda_i - 
	\frac{1}{N} \left( \sum_{i=1}^{k}  i\lambda_i  \right)^2
\end{eqnarray}

Eigenvalues  (\ref{Y}) satisfy the duality 

\begin{eqnarray} \label{dY1}
  	C(D(\lambda, 0), N)=-	C(D(\lambda^T, 0), -N)
\end{eqnarray}

For the proof first note that coefficients at $N$ and $1/N$ terms are the area, and the square of area, respectively, of Young diagram $Y$. Since area doesn't change under transposition, these terms themselves satisfy duality (\ref{dY1}). For other terms, we parameterize $Y$ in a special way, which will make duality clear. 

In (\ref{Y}) Young diagram is parameterized in terms of Dynkin labels $\lambda_i$, which encodes Young diagram having $\lambda_i$ columns of height $i$.  To use this formula both for initial and transposed Young diagrams, we parameterize them in the following way, directly connected with  Dynkin labels of both initial and transposed diagrams. 

For a given Young  diagram let's denote $a_1, a_2,..., a_k$ the number of rows with equal widht, from top to the bottom, and $b_1, b_2,..., b_k$ the number of columns with equal height, from left to right. Clearly $k$ is the same in these two sequences. Below is the example with $k=4$,
and $a_1=1, a_2=1, a_3=2, a_4=3, b_1=1,b_2=2, b_3=2, b_4=1$.

\begin{eqnarray} \label{Ex3}
	\begin{ytableau}
			\none&\none[b_1]&	\none[b_2]&\none&\none[b_3]&\none&\none[b_4]\\
		\none[a_1]&~ & 	~& ~& ~&~&~ \\
		\none[a_2]&~&~&~&~&~   \\
		\none[a_3]&~&~&~\\
	\none &	~&~&~\\
	\none[a_4]&	~ \\
	\none&	~ \\
		\none&~ 
	\end{ytableau}
\end{eqnarray}
 
 Introduce also 
 
 \begin{eqnarray}
 	A_i=a_1+a_2+...+a_i  \,\,\, i=1,2,...,N-1\\
 	B_i=b_1+b_2+...+b_i  \,\,\, i=1,2,...,N-1\\
 	A_0=B_0=0
 \end{eqnarray}

Then  Dynkin labels $(\lambda_1,\lambda_2,...,\lambda_{N-1})$  of given diagram are

\begin{eqnarray}
	(0,...,0,\lambda_{A_1}=b_k, 0,...,0,\lambda_{A_2}=b_{k-1},0,...,...,...,0,\lambda_{A_k}=b_1,0,...,0)
\end{eqnarray}
i.e. among Dynkin labels non-zero are only

\begin{eqnarray}
	\lambda_{A_1}&=&b_k,  \\ \nonumber
	\lambda_{A_2}&=&b_{k-1}, \\   \nonumber
	... \\   \nonumber
	\lambda_{A_k}&=&b_1
\end{eqnarray}

Clearly, transposition of diagram corresponds to interchange $a_i \leftrightarrow b_i$, equivalently $A_i \leftrightarrow B_i$. Particularly, Dynkin labels corresponding to transposed Young diagram are

\begin{eqnarray}
	\lambda_{B_1}&=&a_k,  \\  \nonumber
	\lambda_{B_2}&=&a_{k-1}, \\    \nonumber
	... \\    \nonumber
	\lambda_{B_k}&=&a_1
\end{eqnarray}
and $\lambda_i=0$ for other indexes $i$.

Let's consider first particular case of rectangular Young diagram, which corresponds to the case when non-zero is only one label, say $\lambda_i=\lambda$. Then $a_1=i, b_1=\lambda$. Constant term is $i\lambda^2-i^2\lambda=a_1 b_1^2-b_1 a_1^2$, which is evidently
changing sign under transposition, i.e. under $a_1 \leftrightarrow b_1$.

For the general case consider first part of constant term:

\begin{eqnarray}
	\sum_{i,j=1}^{N-1} min(i,j) \lambda_i \lambda_j=\sum_{a,b=1}^{k} min(A_a,A_b) b_{k+1-a} b_{k+1-b}= \\
	\sum_{a=1}^{k} A_a b_{k+1-a} b_{k+1-a} + 2\sum_{a,b=1, a<b}^{k} A_a  b_{k+1-a} b_{k+1-b}=\\
	\sum_{a=1}^{k} A_a (B_{k+1-a} - B_{k-a})^2  + 2\sum_{a=1}^{k} A_a  (B_{k+1-a}-B_{k-a}) B_{k-a}=\\
	\sum_{a=1}^{k} A_a (B_{k+1-a}^2 - B_{k-a}^2) 
\end{eqnarray}

Second part:

\begin{eqnarray}
	- \sum_{i=1}^{N-1} i^2 \lambda_i = -\sum_{a=1}^k A_a^2 b_{k+1-a} =-\sum_{a=1}^k A_a^2 (B_{k+1-a}-B_{k-a})=\\
	-\sum_{a=1}^{k} B_a (A_{k+1-a}^2-A_{k-a}^2) 
\end{eqnarray}

Alltogether constant term becomes

\begin{eqnarray}
	\sum_{a=1}^{k} A_a (B_{k+1-a}^2 - B_{k-a}^2) 	-\sum_{a=1}^{k} B_a (A_{k+1-a}^2-A_{k-a}^2) 
\end{eqnarray}

This is evidently changing sign under transposition $A_i \leftrightarrow B_i$, so the total Casimir satisfies (\ref{dY1}).

An immediate consequence is that  if diagram is self-dual, i.e. remains unchanged under transposition, the constant term of Casimir is zero.

For the general case one immediately obtain the result (\ref{dY}), since 

\begin{eqnarray}
	C(D(\lambda, \tau), N)=	C(D(\lambda, 0), N)+	C(D(0,\tau), N) + \frac{2}{N} \sum_{i=1}^{N-1} i \lambda_i \sum_{i=1}^{N-1}i\tau_i 
\end{eqnarray}

Last term, which includes product of areas of $\lambda$ and $\tau$ diagrams,  evidently satisfies relation (\ref{dY}), so total expression satisfies, also.

Clearly, if $\lambda$ diagram is transposition of the $\tau$ diagram, the constant term of Casimir is zero.

 \section{Conclusion. Implications for universality}
 
 The problems, discussed in the present paper, arosefrom the universality considerations. Decomposition of powers of adjoint representation of simple Lie algebras into Casimir eigenspaces is suggested in \cite{AIKM,IK25} to be universal. This means that it is invariant w.r.t. the permutations of Vogel's parameters, i.e. for $su(N)$ this implies the duality under $N \leftrightarrow -N$ transformation, both of dimensions of representations, and of Casimir eigenvalues. 
 Particularly, this means that multiplicities of diagrams related by $N \leftrightarrow -N$ should be equal, as observed in known examples.
 Besides that, the theory is invariant under other permutations, e.g. $\beta \leftrightarrow \gamma$, whose manifestation purely in terms of representation theory of $su(N)$ is unclear. 

\section{Acknowledgments}
I'm indebted to S.Krivonos for valuable remarks. 
The work is partially supported by the Science Committee of the Ministry of Science  and Education of the Republic of Armenia under contracts   21AG-1C060 and 24WS-1C031.


\begin{thebibliography}{99}

\bibitem{RP}
R. Penrose, “Applications of negative dimensional tensors,” in Combinatorial mathematics and its applications, D.J.A.Welsh, ed. (Academic Press, New York 1971), 221.

\bibitem{King}
R. C. King {\it The dimensions of irreducible tensor representations of the orthogonal and symplectic groups.} Can. J. Math. {\bf 33} (1972), 176.

\bibitem{PS}
G. Parisi and N. Sourlas {\it Random magnetic fields, supersymmetry and negative
	dimensions.} Phys. Rev. Lett. {\bf 43} (1979), 744.

\bibitem{Cvit} 
P. Cvitanovic and A. D. Kennedy {\it Spinors in negative dimensions.} Phys. Scr. {\bf 26} (1982), 5-12.

\bibitem{Cvitbook}
P. Cvitanovic {\it Group Theory.} Princeton University Press, Princeton, NJ, 2004. 
http://www.nbi.dk/group theory


\bibitem{Mkr} R.L. Mkrtchyan {\it The equivalence of $Sp(2N)$ and $SO(-2N)$ gauge theories.} Physics Letters {\bf 105B} (1981), 174-176.

\bibitem{V0}
P. Vogel,  The Universal Lie algebra, Preprint (1999), https://webusers.imj-prg.fr/\~{}pierre.vogel/grenoble-99b.pdf

\bibitem{MV11}
Ruben L. Mkrtchyan, Alexander P. Veselov, On duality and negative dimensions in the theory of Lie groups and symmetric spaces arXiv:1011.0151, Journal Math. Phys., 52, 083514 (2011), https://doi.org/10.1063/1.3625954

\bibitem{pp}
A.M.Perelomov and V.S.Popov {\it Casimir operators for semisimple Lie groups.}
Mathematics of the USSR-Izvestiya, 2:6 (1968), 1313-1335.


\bibitem{MSV}
R.L. Mkrtchyan, A.N. Sergeev, A.P. Veselov,  Casimir values for universal Lie algebra, arXiv:1105.0115,  Journ. Math.Phys. 53, 102106 (2012).   https://doi.org/10.1063/1.4757763

\bibitem{AM19}	
M.Y. Avetisyan, On Universal Eigenvalues of Casimir Operator,
arXiv:1908.08794, 
Phys. Part. Nucl. Lett. 17(5), pp 779-783 (2020), 
doi:10.1134/S1547477120050039



\bibitem{BR}
A.O. Barut, R. Raczka, Theory of Group Representations and Applications, World Scientific Publ., 1986.

\bibitem{Difr}   di Francesco, P., Mathieu, P. and Senechal, D. (1997) Conformal Field Theory, Springer New York, 1997, 
https://doi.org/10.1007/978-1-4612-2256-9

\bibitem{GG}   
Morikuni Goto and Frank D. Grosshans, Semisimple Lie algebras, Lecture Notes in Pure and Applied Mathematics, V 38, 1978, https://doi.org/10.1201/9781003071778 


\bibitem{MRL25}
R.L.Mkrtchyan, The Casimir eigenvalues on $ad^{\otimes k}$ of SU(N) are linear on N, arXiv:2506.13062, http://arxiv.org/abs/2506.13062,


\bibitem{AIKM} Maneh Avetisyan, Alexey Isaev, Sergey Krivonos and Ruben Mkrtchyan, The uniform structure of $\mathfrak{g}^{\otimes 4}$, Russian Journal of Mathematical Physics, 2024, Vol 31, No 3, pp 379-338, https://doi.org/10.1134/S1061920824030038

\bibitem{IK25} A.P. Isaev and S.O. Krivonos, The split 5-Casimir operator and the structure of $\wedge \mathfrak{ad}^{\otimes 5}$, Izvestiya: Mathematics, 2025, Volume 89, Issue 1, p. 15–25, doi: https://doi.org/10.4213/im9594e

\end{thebibliography}
\end{document}